# The Effects of Superheating Treatment on Distribution of Eutectic Silicon Particles in A357-Continuous Stainless Steel Composite.

**Mazlee Mohd Noor and Shamsul Baharin Jamaludin**

Sustainable Engineering Cluster

Universiti Malaysia Perlis (UniMAP)

01000 Kangar, Perlis, Malaysia

mazlee@unimap.edu.my, sbaharin@unimap.edu.my

**Keywords:** Superheating treatment, eutectic silicon particles, interface, bonding, voids, elemental mapping.

**Abstract** :

In the present study, superheating treatment has been applied on A357 reinforced with 0.5 wt. % (Composite 1) and 1.0 wt.% (Composite 2) continuous stainless steel composite. In Composite 1 the microstructure displayed poor bonding between matrix and reinforcement interface. Poor bonding associated with large voids also can be seen in Composite 1. The results also showed that coarser eutectic silicon (Si) particles were less intensified around the matrix-reinforcement interface. From energy dispersive spectrometry (EDS) elemental mapping, it was clearly shown that the distribution of eutectic Si particles were less concentrated at poor bonding regions associated with large voids. Meanwhile in Composite 2, the microstructure displayed good bonding combined with more concentrated finer eutectic Si particles around the matrix-reinforcement interface. From EDS elemental mapping, it was clearly showed more concentrated of eutectic Si particles were distributed at the good bonding area. The superheating prior to casting has influenced the microstructure and tends to produce finer, rounded and preferred oriented α-Al dendritic structures.

**Introduction**

Microstructure and its formation play a pivotal role in determining relationships between processing and performance of engineering materials [1]. In continuous aluminium matrix composites (AMCs) which are fabricated via casting technique, some of the critical microstructural features are grain size, dendritic arm spacing and silicon morphology in the eutectic phase [2]. Meanwhile in discontinuous AMCs, the most important aspect of the microstructure is the distribution of the reinforcement particles and this depends on the processing and fabrication routes involved [3].

Interfaces constitute an important microstructural feature of composite materials [4]. There are transition zones of finite dimensions at the boundary between the fibre and the matrix where compositional and structural discontinuities can occur over distances varying from an atomic monolayer to over five orders of magnitude in thickness. Most of the mechanical and physical properties of the MMCs such as strength, stiffness, ductility, toughness, fatigue, creep, coefficient of thermal expansion, thermal conductivity and damping properties are dependent on the interfacial characteristic. The interface plays a crucial role in transferring the load efficiently from the matrix to the reinforcement. The strengthening and stiffening of composites are dependent on the load across the interface. A high bonding strength is required at the interface for effective load transfer [5]. A strong bond is usually formed with the reaction between the matrix and the reinforcement meanwhile the reaction products determining the nature of the bond.

A lot of researches have been done regarding eutectic Si particles [6-9]. Meanwhile, the effects of superheating on the refinement of Si phase have been observed in Al-7Si-0.55Mg alloy [10] and hypereutectic Al-Si alloys [11,12]. However, the effects of superheating on the distribution of eutectic Si particles have yet be studied. This is especially at the interface between the matrix alloy and the reinforcement which is a critical region that is affected during the solidification process of metal matrix composites (MMCs). If the interface is not tailored properly, it can lead to the degradation of the properties of the composites [5]. The objective of this research is to study the effects of superheating on the distribution of eutectic silicon particles in A357 continuous stainless steel composites.

**Experimental Procedure**

The raw material used as a matrix alloy in this research work was primary cast ingot A357 alloy. The alloy has been cast by continuous casting process and was delivered in the form of bar. The continuous 304 stainless steel wires (SSw) with 500 µm diameter were used as reinforcement which were embedded in the A357 matrix alloy to produce the composite specimens.

The primary cast ingot A357 alloy was melted in crucibles by using electric furnace. Prior to casting for composite specimens, the 304 stainless steel wires were aligned with uniform dimensions for A357 reinforced with 0.5 and 1.0 wt. % stainless steel composites respectively. The primary cast ingot A357 alloy undergoes superheating treatment at 900ºC for 1 hour before pouring into moulds. The 314 stainless steel mould with aligned 304 stainless steel wires were heated at 250°C for 1 hour. The composite specimens were produced by pouring the molten primary A357 cast ingot into the 314 stainless steel mould at 750°C by gravity casting technique.

**Results and Discussion**

Figure 1 shows the solidification microstructure of Composite 1 which illustrates globular and randomly oriented α-Al dendrites around the continuous stainless steel wire (SSw) reinforcement [13,14]. It also can be seen that lesser amount of eutectic Si particles was present at the grain boundaries in the A357 matrix alloy around the stainless steel wire reinforcement relatively [15]. The microstructure displayed poor bonding which is associated with large voids and less concentrated coarser eutectic Si particles around the matrix-reinforcement interface and also present of macro crack in Composite 1.

Figure 2 shows the close-up of dotted square region in Figure 1. It reveals remarkably poor bonding which again is associated with large voids along the matrix-reinforcement interface. The inexistence of matrix alloy and eutectic Si particles leads to poor bonding and also generation of micro crack along the nearby matrix-eutectic Si interface. However, the nearby acicular eutectics Si particles about more than 20 µm length are remain unaffected.

Figure 3 shows the EDS elemental mapping of the region where poor bonding associated with large voids were seen in Composite 1. It shows the intensity of various elements (Al, Si and Mg) at and around the poor bonding region in this composite. From EDS elemental mapping, it was clearly showed less concentrated presence of eutectic Si particles at the poor bonding region which are associated with large voids. However, more concentrated of eutectic Si particles can be observed at the good bonding region (rounded dotted line) as shown in Si element (Figure 1). Away from the poor bonding region, the eutectic Si particles are found to be evenly distributed throughout the A357 matrix alloy [16].

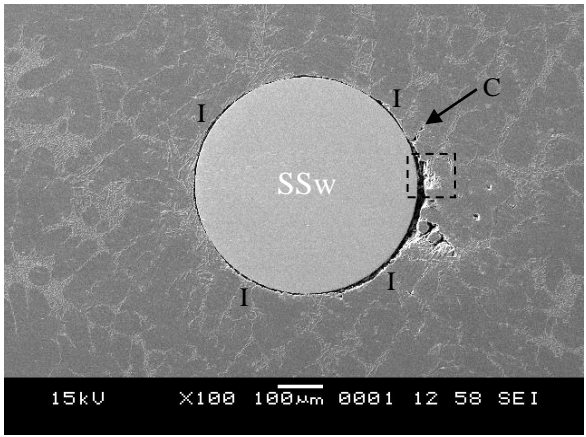

Figure 1 SEM micrograph of Composite 1. Dotted rectangular region shows poor bonding associated with large voids. Poor bonding denoted by I and macro crack denoted by C.

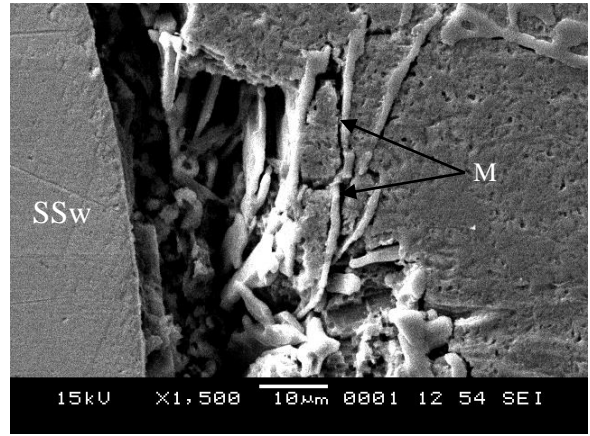

Figure 2 Close-up of matrix-reinforcement interface Composite 1 (dotted rectangular region in Figure 1). Micro cracks denoted by M.

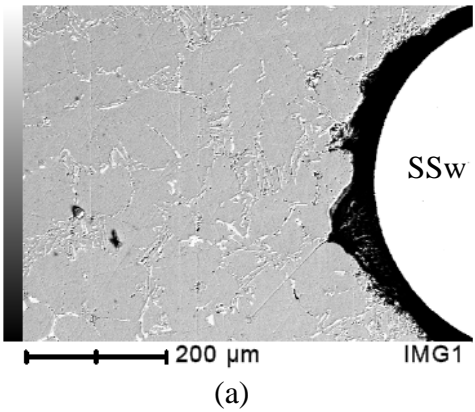

(a)

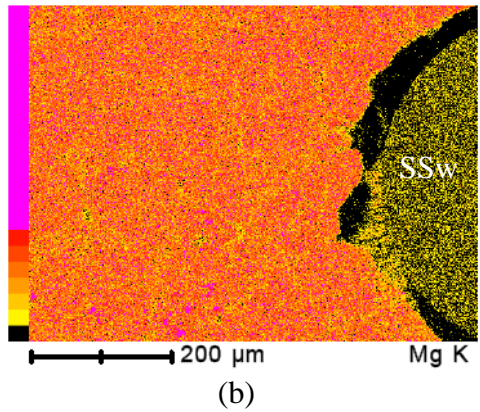

(b)

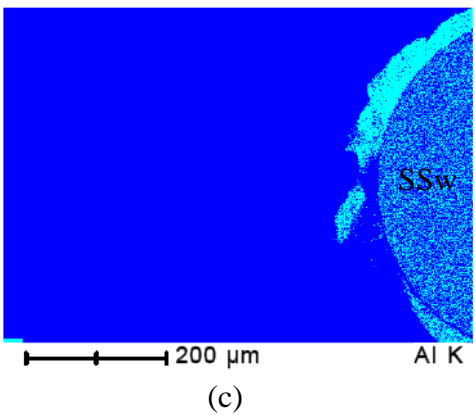

(c)

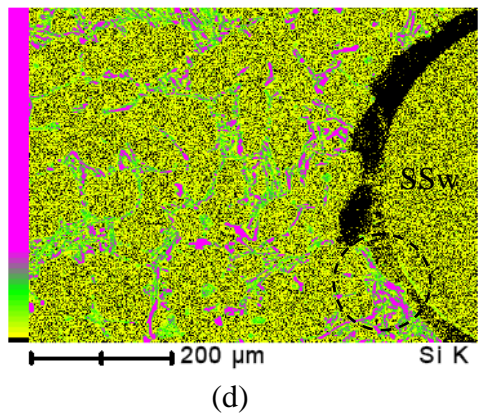

(d)

Figure 3 EDS elemental mapping of poor bonding associated with large voids in Composite 1; (a) Secondary electron image, (b) Mg, (c) Al, (d) Si. Rounded dotted line shows nearly free poor bonding region in (d).

Figure 4 shows the solidification microstructure of Composite 2 which also illustrates globular and randomly oriented α-Al dendrites around the continuous stainless steel wires reinforcement. Heavily segregated of finer eutectic Si particles also can be observed at the grain boundaries in the A357 matrix alloy around the stainless steel wire reinforcement. The microstructure displayed good bonding combined with more concentrated finer eutectic Si particles around the matrix-reinforcement interface.

Preferential segregation of finer acicular eutectic Si particles about 10-15 μm length can be found close to the good bonding (N1) at the matrix-reinforcement interface in Figure 5. The superheating treatment prior to casting has influenced the microstructure and tends to refine the eutectic Si particles and change the shape of α-Al dendrite from coarser, irregular and non-equiaxed to finer, rounded and preferred oriented α-Al dendritic structures. This finding was similar to what was reported by Wang et al. [17] where most of α-Al dendrites were broken to rounded and smaller structures due to superheating treatment at 840ºC for 20 minutes in A356 alloys. Meanwhile, the entire α-Al dendrites can be changed to rounded and smaller structures as reported by Wanqi et al. [10] by superheating the Al-7Si-0.55Mg alloy at 900ºC for 30 minutes. In addition, in this research, the shape of α-Al dendrites also has changed to preferred oriented structure. This change may be attributed to the longer soaking time of 1 hour at superheating temperature of 900ºC.

In terms of refinement of eutectic Si particles, Wanqi et al. [10] reported that if the superheat temperature is kept below 800ºC, there was no obvious effect on the refinement of the Si particles in A356 alloy. However, over 800ºC, the Si particles were significantly refined. For example, a superheating process above 850ºC for 30 minutes has modified the interdendritic eutectic Si particles and was comparable to that achieved by Sr addition. Specifically, at superheating temperature of 900ºC, the refinement of the Si particles by superheating is nearly comparable to that obtained with 0.01 percent Sr addition [10].

Refinement of the Si particles by superheating was influenced by the existence of Mg in the A356 alloy, since Wanqi et al. [10] did not find any such effect in Al-Si binary alloys which containing no Mg. The superheating of the liquid metals modifies the Si phase through two mechanisms. The first one is to reduce the possibility of the heterogeneous nucleation of Si phase. The second is to homogenise Mg distribution so that it will influence the growth of Si phase.

Figure 6 shows the EDS elemental mapping of good bonding in Composite 2. It shows the intensity of various elements (Al, Si and Mg) at and around the good bonding region in this composite. From the EDS elemental mapping, it was clearly showed more concentrated presence of eutectic Si particles at the good bonding region. Away from the good bonding region, the eutectic Si particles are found to be evenly distributed throughout the A357 matrix alloy.

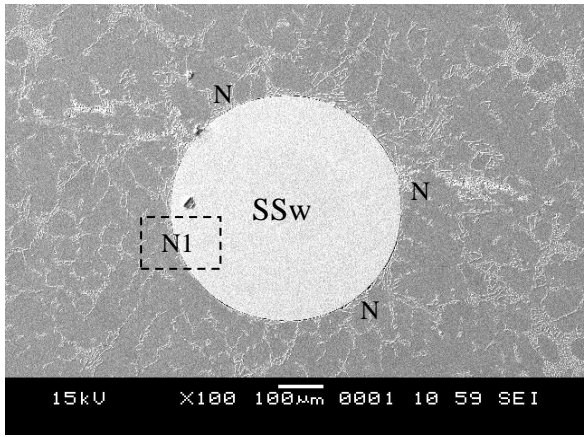

Figure 4 SEM micrograph of Composite 2. Good bonding denoted by N.

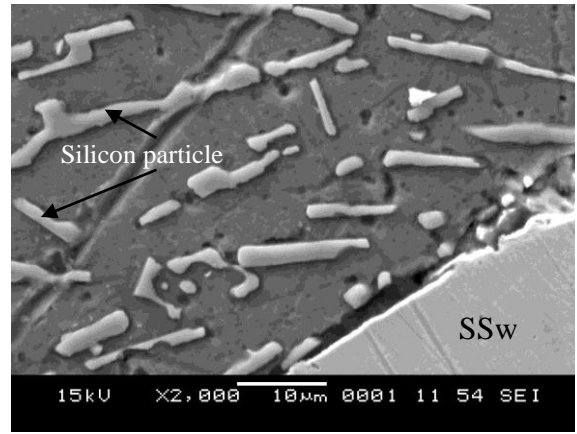

Figure 5 Close-up of good bonding interface in Composite 2 (dotted rectangular region, N1 in Figure 4).

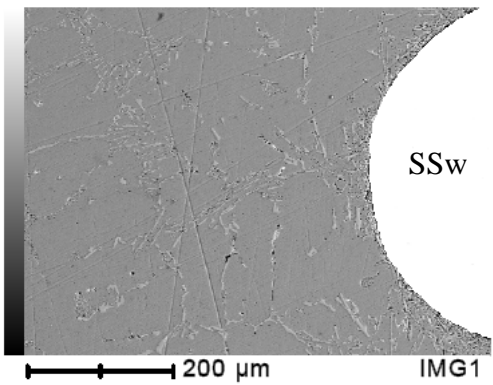

(a)

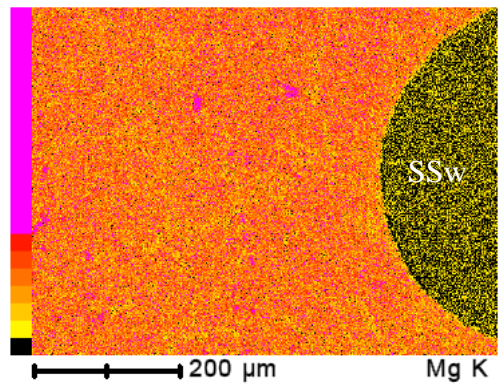

(b)

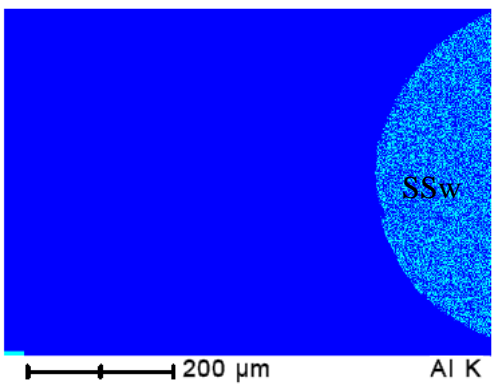

(c)

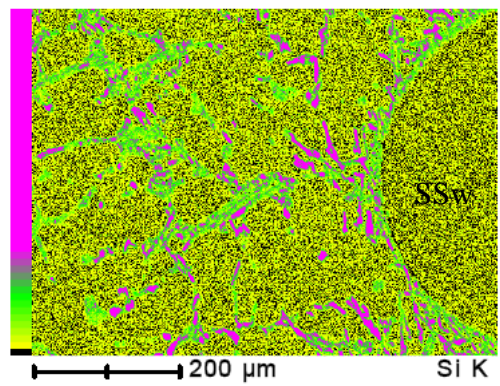

(d)

Figure 6 Elemental EDS mapping of good bonding in Composite 2; (a) Secondary electron image, (b) Mg, (c) Al, (d) Si.

# Conclusions

The following conclusions can be drawn from this study:

1) In Composite 1, the effects of superheating treatment have resulted less concentrated presence of eutectic Si particles at the poor bonding region which are associated with large voids. More concentrated of eutectic Si particles distribution can be observed at the nearly free poor bonding region.

2) In Composite 2, the effects of superheating treatment have resulted more concentrated presence of eutectic Si particles at the good bonding region.

# Acknowledgement

The authors is grateful to UniMAP for the financial support under research grant (9014-00008).